\documentclass[conference]{IEEEtran}
\IEEEoverridecommandlockouts
\usepackage{cite}
\usepackage{algorithmic}
\usepackage{array}
\usepackage{graphicx}
\usepackage{subcaption}
\usepackage{textcomp}
\usepackage{float}
\usepackage{balance}
\usepackage{multirow}
\usepackage{xspace}
\usepackage[htt]{hyphenat}
\usepackage{xcolor}
\usepackage[most]{tcolorbox}
\usepackage{booktabs}
\usepackage{listings}
\usepackage{tikz}
\usepackage{makecell}
\usepackage{subcaption}
\usepackage{ragged2e}                       
\usepackage{booktabs, makecell, multirow, tabularx}
\PassOptionsToPackage{hyphens}{url}\usepackage{hyperref}

\newcommand*\circled[1]{\tikz[baseline=(char.base)]{
            \node[shape=circle,draw,inner sep=0.7pt,fill=white,text=black] (char) {\textbf{\texttt{#1}}};}}

\newcommand{\visibility}{\textcolor{black}{visibility}\xspace}
\newcommand{\informativity}{\textcolor{black}{informativity}\xspace}

\newcommand{\name}{\textsc{\textcolor{black}{Installamatic}}\xspace}

\def\BibTeX{{\rm B\kern-.05em{\sc i\kern-.025em b}\kern-.08em
    T\kern-.1667em\lower.7ex\hbox{E}\kern-.125emX}}

\makeatletter
\newcommand\notsotiny{\@setfontsize\notsotiny\@vipt\@viipt}
\makeatother

\begin{document}
\bstctlcite{IEEEexample:BSTcontrol}

\title{
Beyond pip install: Evaluating LLM Agents for the Automated Installation of Python Projects
}

\author{\IEEEauthorblockN{Louis Milliken}
\IEEEauthorblockA{\textit{School of Computing} \\
\textit{KAIST}\\
Daejeon, Republic of Korea \\
lmilliken@kaist.ac.kr}
\and
\IEEEauthorblockN{Sungmin Kang}
\IEEEauthorblockA{\textit{School of Computing} \\
\textit{KAIST}\\
Daejeon, Republic of Korea \\
sungmin.kang@kaist.ac.kr}
\and
\IEEEauthorblockN{Shin Yoo}
\IEEEauthorblockA{\textit{School of Computing} \\
\textit{KAIST}\\
Daejeon, Republic of Korea \\
shin.yoo@kaist.ac.kr}
}

\maketitle

\begin{abstract}
    Many works have recently proposed the use of Large Language Model (LLM) based agents for performing `repository level' tasks, loosely defined as a set of tasks whose scopes are greater than a single file.
    This has led to speculation that the orchestration of these repository-level tasks could lead to software engineering agents capable of performing almost independently of human intervention. However,
    of the suite of tasks that would need to be performed by this autonomous software engineering agent, we argue that one important task is missing, which is to fulfil project level dependency by installing other repositories.
    To investigate the feasibility of this repository level installation task,
    we introduce a benchmark of of repository installation tasks curated from 40 open source Python projects, which includes a ground truth installation process for each target repository.
    Further, we propose \name, an agent which aims to perform and verify the installation of a given repository by searching for relevant instructions from documentation in the repository. 
    Empirical experiments reveal that that 55\% of the studied repositories can be automatically installed by our agent at least one out of ten times.
    Through further analysis, we identify the common causes for our agent's inability to install a repository,  discuss the challenges faced in the design and implementation of such an agent and consider the implications that such an agent could have for developers.
\end{abstract}


\section{Introduction}
\label{sec:intro}

Large Language Models (LLMs) are statistical language models, typically based on the Transformer~\cite{Vaswani2017aa} Deep Neural Network architecture, that are trained on large amounts of data with the goal of predicting the next token in a sequence of text.
LLMs trained with large corpora have shown emergent capabilities~\cite{Wei2022ab}, including in-context learning~\cite{Brown2020aa},
i.e., the ability to perform tasks for which the models have not been explicitly trained.
In particular, LLMs are capable of software related tasks when the training corpora include source code~\cite{Chen2021ec}. 

LLMs have been rapidly adopted into software engineering~\cite{Fan2023yu}.
Initially, the target applications were of small, local scopes:
synthesizing individual functions~\cite{Chen2021ec}, mutating inputs for fuzzing~\cite{Xia2024aa}, generating a unit test case~\cite{Lemieux2023aa}, and generating single-line patches~\cite{Xia2022aa}, etc.
As increasingly larger models are being developed and made available~\cite{naveed2023comprehensive},
more advanced prompting techniques (such as Chain-of-Thoughts~\cite{wei2022chain}, ReAct~\cite{Yao2022qf},and self-consistency~\cite{Wang2023aa})
and LLM-based architectures (such as agent~\cite{kang2024quantitative} and multi-agent architectures~\cite{Yoon2024aa,tao2024magis}) have been adopted to perform ``repository-level'' tasks\cite{bairi2024codeplan,wang2024teaching}, which we define in this paper to be tasks that require reading and/or writing multiple files in a given repository.

We argue that all of these repository level agents almost exclusively focus on \emph{code management} tasks, i.e., tasks that analyse or manipulate the source code of a repository.
On the other hand, developers also frequently deal with environment management, for which LLM based agents have not been applied to, to the best of our knowledge. 

Thus, to fully understand how LLM based agents could aid developers in practice, it is imperative to investigate their ability to perform \emph{environment management} tasks.
To this end, this paper presents a novel task for LLM-based agents, which is to install a given code repository, as well as to validate the installation.
Despite it being a common task for many developers, attempts to automate the installation of open-source projects have not been made in previous works.
Dagenais et al. \cite{dagenais2010moving} found that inadequate deleloper documentation is an obstacle for new newcomers to a software project.
Moreover, in a survey by Aghajani et al.\cite{aghajani2020software}, 68\% of developers asked said that incomplete documentation of the installation, deployment and release of a project is an important issue,
and 63\% claimed that inappropriate installation instructions were a common issue as well.
As such, we believe that a tool that can be used to attempt automatic installation could lessen developer frustration and improve productivity.




\begin{figure*}[t]
    \centering
        \begin{subfigure}[T]{0.32\textwidth}
            \centerline{\includegraphics[width=\linewidth]{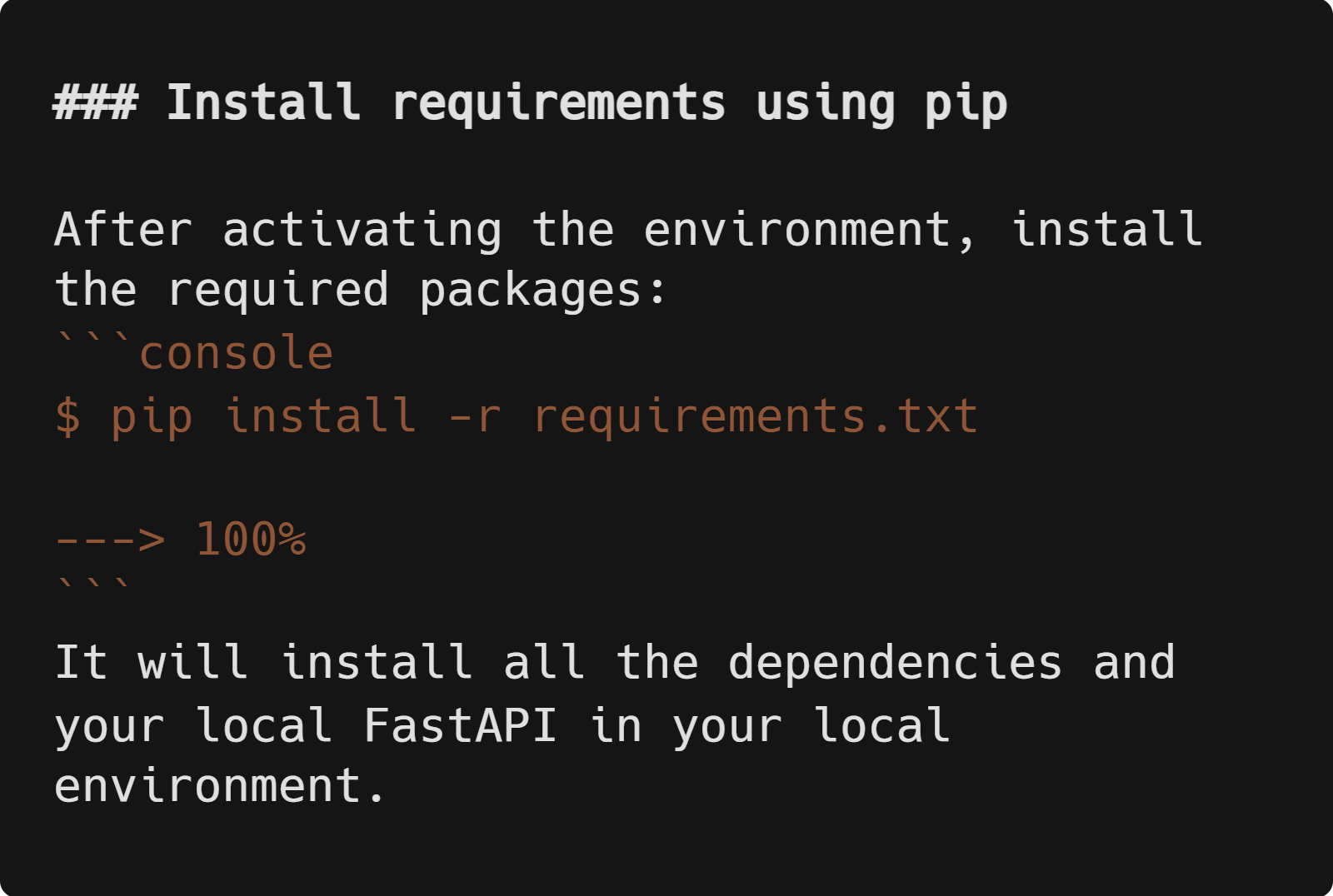}}
            \caption{Example of an install-relevant piece of documentation.}
            \label{fig:FastAPI-relevant}
        \end{subfigure}
    \hfill
        \begin{subfigure}[T]{0.32\textwidth}
            \centerline{\includegraphics[width=\linewidth]{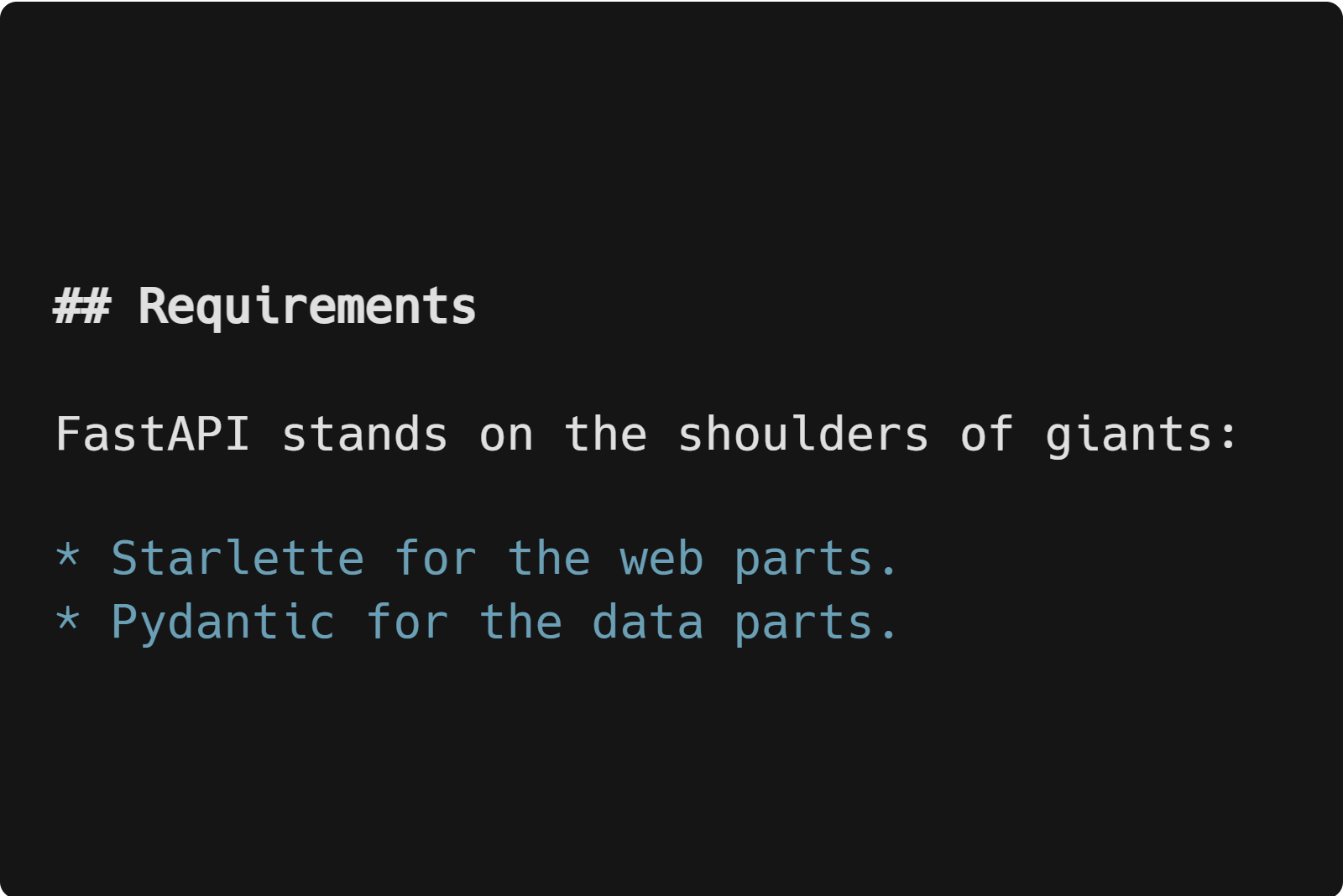}}
            \caption{Example of a non install-relevant piece of documentation. (lightly edited for clarity)}
            \label{fig:FastAPI-nonrelevant}
        \end{subfigure}
    \hfill
        \begin{subfigure}[T]{0.3\textwidth}
            \centerline{\includegraphics[width=\linewidth]{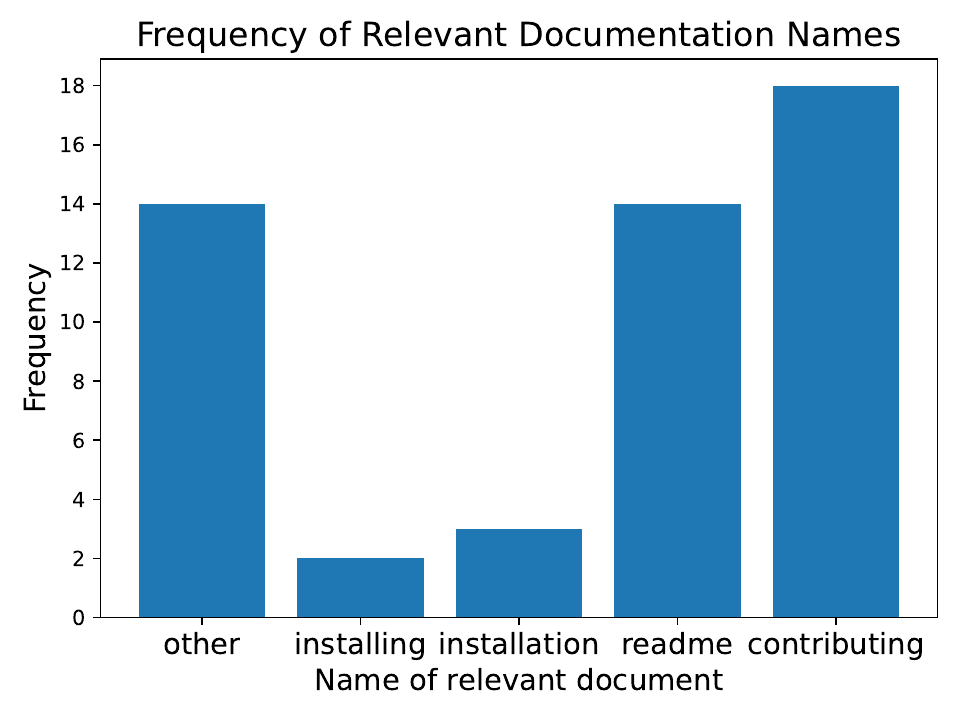}}
            \vspace{-5pt}
            \caption{Frequencies of install-related document filenames}
            \label{fig:doc-freq}
        \end{subfigure}
    \caption{Inspection of repository contents}
\end{figure*}

In order to evaluate LLM-based agents for this task, we create a dataset of 40 open source Python repositories to serve as a benchmark for the performance of environment management agents.
Each repository is assigned a set of tags indicating its installation method and a ground truth working installation file. To ensure the safe execution of LLM generated installation scripts, we also provide an interface for connecting the LLM agent to a virtual machine in which to safely attempt installation. 
We make our research artefact, including this dataset, publicly available to facilitate the continued evaluation of LLM based agents: \url{https://github.com/coinse/installamatic}.

Using this dataset, we investigate how well an LLM-based agent can resolve the installation task by creating \name, which attempts to automatically install and test repositories based on their contents.
Given a target repository, the agent makes use of search tools to navigate through its contents, and inspect files to find information relevant to the installation or testing process.
Once the agent has completed its documentation gathering, we prompt it to write a Dockerfile that, when placed inside the target repository,
installs any required dependencies and runs the test suite of the target repository to confirm the success of the installation.

Our empirical evaluation of \name's ability to perform the installation task
shows that our LLM-based agent can successfully install 21 out of 40 repositories, achieving the success rate of 55\%.
Through further experimentation, we present lessons for designers of future environment management agents, as well as open source project maintainers.
For future development of environment management agents, we suggest the inclusion of a repair step after an agent's initial installation attempt.
For maintainers of open source repositories, we suggest providing code examples of the installation process when writing installation instructions in the repository's documentation.
We also identify several challenges specific to environment manipulation tasks, such as gathering task-relevant information from the repository.

The technical contributions of this paper are:

\begin{itemize}
    \item A dataset of 40 open source Python repositories, designed to serve as a benchmark for evaluating the effectiveness of repository-level agents' understanding of repository contents and environment management tasks.
    \item The initial design of a repository-level agent, \name, capable of searching for and reading documentation, then writing a Dockerfile to install and test the repository based on the gathered information.
    \item An outline of the key challenges that future researchers will likely face when developing a repository-level agent for documentation.
\end{itemize}

The remainder of the paper is structured as follows.
Section~\ref{sec:benchmark-construction} describes the benchmark construction, analysis and labelling process; Section~\ref{sec:auto-eval} provides a breakdown of our proposed agent, \name; Section~\ref{sec:experimental-setup} presents the three research questions we answer, as well as define key metrics in our analysis; Section~\ref{sec:results} examines the results of our experiments; Section~\ref{sec:related-work} compares our work with related literature; Section~\ref{sec:validity} discusses the threats to validity; and Section~\ref{sec:conclusion} concludes.








\section{Dataset Construction}
\label{sec:benchmark-construction}


In order to create a benchmark for the task of automatic installation,
we collect and present a dataset of 40 open source Python repositories from GitHub and provide the correct method of installation,
as well as the location of any relevant documentation, for each repository.
In cases where no relevant documentation was found,
the appropriate installation method was identified through inspection of non-documentation files in the repository and manual attempts to install the repository. 

Repositories were sampled using the GitHub API from several different ranges of stars: 1k-5k, 5k-10k, 10k-20k and \textgreater 20k. From each range, the 10 most recently updated repositories at the time of collection were chosen, meaning that all repositories in the dataset have been in active development until at least August of 2024. The dataset contains the commit ID that corresponds to the time of collection.

We only choose repositories with test suites located in a \texttt{test} or \texttt{tests} directory, to ensure that we have a consistent oracle to determine the success of the installation process. We argue that successful executions of test suites can serve as objective and automatable evidence for successful installations. While this decision was necessary to evaluate the capability of our agent, it does bias our sample towards the repositories with test suites, an issue that is discussed further in Section~\ref{sec:RQ3}. Table~\ref{tab:repos} shows the collected repositories.

\begin{table*}[th]
\centering
\caption{List of repositories in the dataset\label{tab:repos}}
\scalebox{0.9}{
\begin{tabular}{lll|lll}
\toprule
$\star$                                                             & Name                & URL                                                      & $\star$                                                              & Name               & URL \\ \midrule
\parbox[t]{2mm}{\multirow{10}{=}{\rotatebox[origin=c]{90}{1k-5K}}}  & icloud-drive-docker & \url{https://github.com/mandarons/icloud-drive-docker}   & \parbox[t]{2mm}{\multirow{10}{=}{\rotatebox[origin=c]{90}{10k-20K}}} & yfinance           & \url{https://github.com/ranaroussi/yfinance}  \\
                                                                    & django-stubs        & \url{https://github.com/typeddjango/django-stubs}        &                                                                      & beets              & \url{https://github.com/beetbox/beets}  \\
                                                                    & pennylane           & \url{https://github.com/PennyLaneAI/pennylane}           &                                                                      & starlette          & \url{https://github.com/encode/starlette}  \\
                                                                    & X-AnyLabeling       & \url{https://github.com/CVHub520/X-AnyLabeling}          &                                                                      & datasets           & \url{https://github.com/huggingface/datasets}  \\
                                                                    & opencompass         & \url{https://github.com/open-compass/opencompass}        &                                                                      & mypy               & \url{https://github.com/python/mypy}  \\
                                                                    & R2R                 & \url{https://github.com/SciPhi-AI/R2R}                   &                                                                      & sympy              & \url{https://github.com/sympy/sympy}  \\
                                                                    & Torch-Pruning       & \url{https://github.com/VainF/Torch-Pruning}             &                                                                      & ydata-profiling    & \url{https://github.com/ydataai/ydata-profiling}  \\
                                                                    & scvi-tools          & \url{https://github.com/scverse/scvi-tools}              &                                                                      & spotify-downloader & \url{https://github.com/spotDL/spotify-downloader}  \\
                                                                    & sabnzbd             & \url{https://github.com/sabnzbd/sabnzbd}                 &                                                                      & qlib               & \url{https://github.com/microsoft/qlib}  \\
                                                                    & dlt                 & \url{https://github.com/dlt-hub/dlt}                     &                                                                      & scapy              & \url{https://github.com/secdev/scapy} \\ \midrule
\parbox[t]{2mm}{\multirow{10}{=}{\rotatebox[origin=c]{90}{5k-10K}}} & camel               & \url{https://github.com/camel-ai/camel}                  & \parbox[t]{2mm}{\multirow{10}{=}{\rotatebox[origin=c]{90}{20K+}}}    & fastapi            & \url{https://github.com/tiangolo/fastapi}  \\
                                                                    & boto3               & \url{https://github.com/boto/boto3}                      &                                                                      & black              & \url{https://github.com/psf/black}  \\
                                                                    & cloud-custodian     & \url{https://github.com/cloud-custodian/cloud-custodian} &                                                                      & tqdm               & \url{https://github.com/tqdm/tqdm}  \\
                                                                    & aim                 & \url{https://github.com/aimhubio/aim}                    &                                                                      & rich               & \url{https://github.com/Textualize/rich}  \\
                                                                    & speechbrain         & \url{https://github.com/speechbrain/speechbrain}         &                                                                      & open-interpreter   & \url{https://github.com/OpenInterpreter/open-interpreter}  \\
                                                                    & nonebot2            & \url{https://github.com/nonebot/nonebot2}                &                                                                      & core               & \url{https://github.com/home-assistant/core}  \\
                                                                    & moto                & \url{https://github.com/getmoto/moto}                    &                                                                      & sherlock           & \url{https://github.com/sherlock-project/sherlock}  \\
                                                                    & instructor          & \url{https://github.com/jxnl/instructor}                 &                                                                      & spaCy              & \url{https://github.com/explosion/spaCy}  \\
                                                                    & numba               & \url{https://github.com/numba/numba}                     &                                                                      & you-get            & \url{https://github.com/soimort/you-get}  \\
                                                                    & pymc                & \url{https://github.com/pymc-devs/pymc}                  &                                                                      & textual            & \url{https://github.com/Textualize/textual}  \\
\bottomrule
\end{tabular}}
\end{table*}

Each repository has been manually inspected to produce three different types of metadata: the aforementioned list of all install-relevant documents, an exemplar Dockerfile that successfully installs and tests the repository, and a set of tags indicating the expected installation and testing methods.
The Dockerfile writing and tag assignment process is described in more detail in \ref{sec:install-methods}.

\subsection{Documentation Structure of Open Source Python Projects}
\label{sec:doc_structure}

We define a document to be `install-relevant' if it makes explicit references to the process of installing the target repository's dependencies and executing its test suite. For example, Figure~\ref{fig:FastAPI-relevant} is an example of install-relevant documentation, as it clearly shows one of the the commands needed to be run to set up a development environment. On the other hand, Figure~\ref{fig:FastAPI-nonrelevant} shows an example of a non install-relevant piece of documentation from the \texttt{README.md} file of the FastAPI repository: this is not install-relevant as it does not affect the installation process for a developer.


After identifying the install-relevant documents for each repository, we have found 29 unique file paths leading to install-relevant documentation, and 18 different names for files containing install-relevant documentation, ignoring differences between file type, case sensitivity and the use of `\texttt{-}' and `\texttt{\_}'. Figure~\ref{fig:doc-freq} shows the distribution of documentation file names. While one may expect the `readme' file of a repository to contain information relevant to the installation of a repository, we find that this is often not the case.

Names `contributing' and `readme' are considerably more common than any other file name: files with these names contain install-relevant information in 40\% and 35\% of repositories in the dataset, respectively.
After these two, there are no other file names which are install-relevant and occur more than three times.
Note that the purpose of a `contributing' file is to instruct developers new to the project how to go about making contributions to the project. 
As such, instructions on dependency management, environment setup and testing are commonplace in files named `contributing'.


It is worth noting that many projects host much of their documentation on external websites, though occasionally the source files for these external websites are stored in the repository itself.
In such cases, the documentation will be often stored inside a \texttt{docs} directory, meaning that it is still accessible in the repository.
However, it is considerably less visible and consequently harder to find, compared to documentation stored in the repository's root directory.
The diversity in possible locations for install-relevant documentation shows that there is no agreed upon structure for documentation of Python repositories.

\subsection{Methods of Installation and Testing} 
\label{sec:install-methods}

In addition to identifying the locations of install-relevant documentation,
our dataset contains a Dockerfile that would, when placed inside of the target repository,
install any dependencies, and run the test suite to confirm their successful installation.
During this process, we create a series of coarse-grained categories for the different types of commands used during installation.
Table~\ref{tab:tags} lists the resulting 17 tags.

\lstset{
    basicstyle=\ttfamily\notsotiny
}

\begin{table*}[th]
\centering
\caption{List of Installation Tags}
\label{tab:tags}
\scalebox{0.85}{
\begin{tabular}{ll|ll}
\toprule
\thead[l]{Tag} & \thead[l]{Description} & \thead[l]{Tag} & \thead[l]{Description} \\ \midrule
\textbf{requirements}       & \makecell[l]{Installation of dependencies using \\\lstinline+pip install -r requirements.txt+.}                                                                       & \textbf{install-other}      & \makecell[l]{Installation of dependencies through means not listed\\above, such as a custom script contained in the repository} \\ \midrule
\textbf{requirements-extra} & \makecell[l]{Installation of dependencies from additional\\requirements files, such as \lstinline+requirements-test.txt+.}                                            & \textbf{pytest}       & \makecell[l]{Tests are run using pytest.} \\ \midrule
\textbf{pip-extra}          & \makecell[l]{Requiring the installation of something other\\than Poetry or the contents of a requirements file.}                                                      & \textbf{pytest-extra} & \makecell[l]{Additional arguments need to be provided to pytest,\\such as specifying the location of the tests or additional flags.} \\ \midrule
\textbf{Poetry}             & \makecell[l]{Installation of dependencies using the Poetry\\dependency manger.\footnote{https://python-poetry.org/}}                                                  & \textbf{tox}          & \makecell[l]{Tests are run using Tox.} \\ \midrule
\textbf{Poetry-extra}       & \makecell[l]{Installation of dependencies using Poetry, with\\additional arguments, e.g. \lstinline+poetry install+\\\lstinline+--no-interaction --with sentry-sdk+.} & \textbf{unittest}     & \makecell[l]{Tests are run using Python's built in unittest\\command.} \\ \midrule
\textbf{make-install}       & \makecell[l]{Installation of dependencies using a makefile,\\typically commands such as \lstinline+make install+ or \lstinline+make init+.}                           & \textbf{make-test}    & \makecell[l]{Tests are run using a makefile with a command such\\as \lstinline+make test+.} \\ \midrule
\textbf{install-self}       & \makecell[l]{The project itself needs to be installed in the\\working environment, e.g. by running \lstinline+pip install -e .+}                                      & \textbf{test-other}   & \makecell[l]{Tests are run some other way, such as a \lstinline+test.py+ file.} \\ \midrule
\textbf{install-pytest}     & \makecell[l]{The Pytest\footnote{https://docs.pytest.org/en/stable/} library needs to be installed\\manually.}                                                        & \textbf{bash-extra}   & \makecell[l]{Requiring additional bash commands to set up\\the repository, such as creating new directories\\or granting permissions to certain files.} \\ \midrule                      & \\ 
\textbf{install-tox}        & \makecell[l]{The Tox\footnote{https://tox.wikinstalli/en/4.17.1/} library needs to be installed manually.}                                                            &                       & \\ 
\bottomrule
\end{tabular}}
\end{table*}

After assigning appropriate tags to each of the 40 repositories, we are left with 31 unique combinations of tags, meaning that there are 31 different methods of installing the dependencies and running the tests of the Python repositories we sampled.
Such a wide variety of installation methods leads us to consider the use of Large Language Models (LLMs) as a means of automatic installation;
their in-context learning capability~\cite{brown2020languagemodelsfewshotlearners} makes them an ideal candidate in tasks like this, where the expected output 
(in this case, a working Dockerfile) can vary considerably, and is highly dependant on the contents the documentation in the repositories.

\section{\name: Automatic Installation Agent} 
\label{sec:auto-eval}

This section presents \name, an LLM-based agent that attempts to automatically install a given open source Python project. \name performs its task in two stages.
First, it gathers documentation related to the installation.
Second, it tries to build and repair a Dockerfile\footnote{https://www.docker.com/} to install and test the target repository.
This section first explains how \name searches for files relevant to installation process, and subsequently describes the two stages of its task.


\subsection{Repository Search Process}
\label{sec:repo_search}

\begin{figure*}
    \centering
        \begin{subfigure}[T]{0.49\textwidth}
            \centerline{\includegraphics[width=\linewidth]{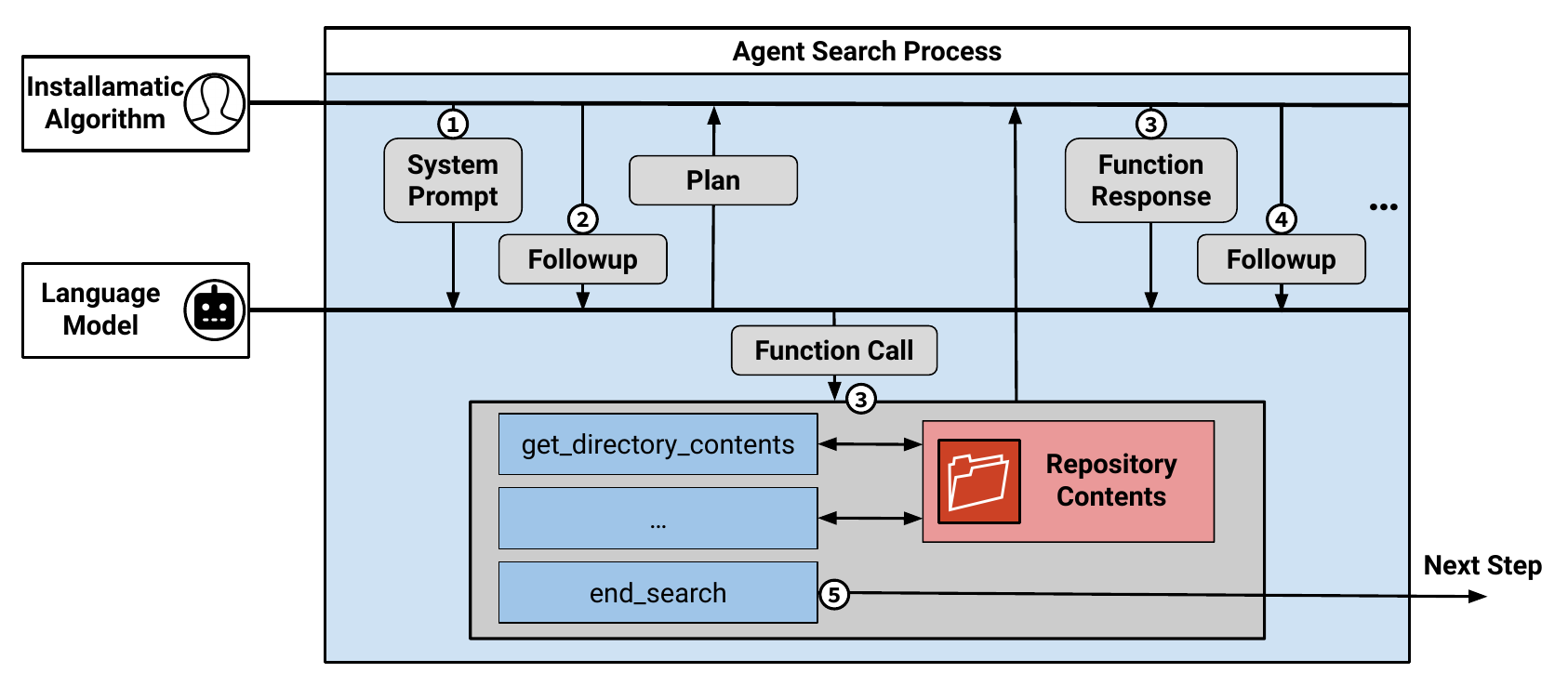}}
            \caption{Diagram of the agent's search process}
            \label{fig:tool-loop}
        \end{subfigure}
    \hfill
        \begin{subfigure}[T]{0.49\textwidth}
            \centerline{\includegraphics[width=\linewidth]{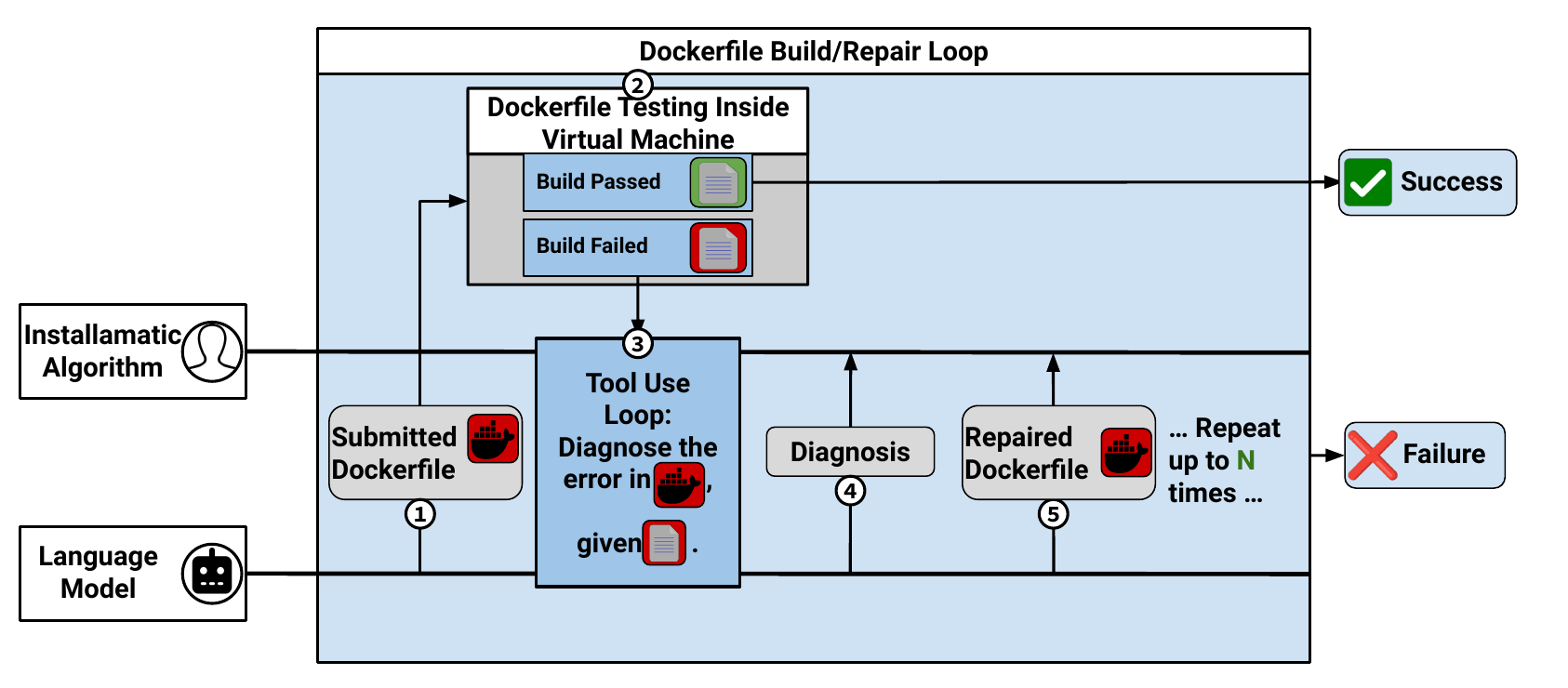}}
            \caption{Agent control during Dockerfile test/repair step}
            \label{fig:dockerfile-repair}
        \end{subfigure}
    \hfill
    \caption{Diagrams of \name's processes}
    \vspace{-15pt}
\end{figure*}

\name needs to search through the contents of the repository during both the documentation gathering and the Dockerfile build/repair step, although, its exact goal can differ in these steps.
To this end, we equip \name with a generic method to search through the contents of the repository, which can be reused in different stages.
An overview of this search method is shown in Figure ~\ref{fig:tool-loop}.
Inspired by previous works finding benefits such as improved performance~\cite{shinn2023reflexionlanguageagentsverbal} 
and explainability~\cite{kang2023explainableautomateddebugginglarge} by emulating human behaviour, we chose to use an LLM-guided search step over traditional search methods, 
such as BM25~\cite{BM25} or neural embeddings~\cite{mitra2017neural}.

We include the contents of the target repository's root directory in the prompt at the start of the search process, then provide several tools that enable the agent to further navigate through the contents of the repository. The four basic functions we provide to \name for navigation are as follows:

\begin{itemize}
    \item \textbf{get\_directory\_contents}: Given a path to a directory the agent has already seen, returns the names of all files and sub-directories within that directory.
    \item \textbf{get\_file\_contents}: Given a path to a file, returns its contents. If the file is human readable and structured, such as a \texttt{.md} or a \texttt{.rst} file, then the section headers are extracted and returned instead.
    \item \textbf{inspect\_header}: Given a path to a file and a name of a section in that file, returns the contents of that section. This is used to minimize the amount of distracting contents shown to the agent.
    \item \textbf{check\_presence}: Given a filename, checks whether it exists. This is provided as a sanity check, to prevent the agent from hallucinating files that are not there.
\end{itemize}

In addition to these functions that are provided during search, \name can access additional functions that are available for specific search tasks it is performing. Examples of these specific tasks will be provided below. All prompts used in this process are listed in the appendix, available in the research artifact.

To start the search process, the agent is shown a system prompt explaining its task and providing the contents of the target repository's root directory (Fig.~\ref{fig:tool-loop}\circled{1}).
The agent then enters a search loop, starting with being sent a query asking the agent to plan its next move in natural language,
followed by a second query offering the agent tools with which it will carry out its plan (Fig.~\ref{fig:tool-loop}\circled{2}).
This two step approach is designed both to improve the reasoning ability of the agent, inspired by previous work on chain of thought reasoning~\cite{wei2022chain}, 
and to provide a better, more understandable explanation of the agent's behaviour when examining the results.
After these two steps, a function is chosen based on the agent's response, and executed to return the relevant result to the agent (Fig.~\ref{fig:tool-loop}\circled{3}). Once the result of the function has been sent to the agent, the prompt to make the agent plan its next move is sent again (Fig.~\ref{fig:tool-loop}\circled{4}). 
This process is repeated until the agent deems the search to have ended (Fig.~\ref{fig:tool-loop}\circled{5}). Additionally, the system (Fig.~\ref{fig:tool-loop}\circled{1}), follow-up (Fig.~\ref{fig:tool-loop}\circled{2})), and function response prompts (Fig.~\ref{fig:tool-loop}\circled{3}), are changed depending on the specific task the agent is currently performing
For example, the follow-up prompt during the documentation gathering step will remind the agent to use the provided tool for recording documentation whenever it identifies a document as install-relevant.
In contrast, when performing a diagnosis of an error during the dockerfile repair process, the follow-up prompt will instead remind the agent to stop searching through the documentation once it has gathered sufficient information to suggest a fix to the Dockerfile. By adopting this modular approach, we are able to clearly define different states in which the agent is operating, which has been shown to aid the agent's decision making when given a complex task\cite{bouzenia2024repairagentautonomousllmbasedagent}.


\subsection{Documentation Gathering Step}

During the first stage, \name is tasked with searching through the repository and identifying any files that it considers to have information relevant to either the installation or testing process. In addition to the four basic functions, the agent is given more tools: \textbf{submit\_documentation}, which records a document as being install-relevant, and \textbf{finished\_search}, which signals the end of this stage and the beginning of the next step.

\subsection{Dockerfile Build/Repair Step}
After documentation gathering step, we task the agent to summarise the gathered information in natural language. \name is once again given access to the basic search functions. In this stage, it can only access the files that it previously selected as being install-relevant. Once it has finished the search process and used the \textbf{submit\_summary} tool to give its summary, it is prompted to write a Dockerfile to install dependencies and run tests. 


Figure~\ref{fig:dockerfile-repair} illustrates the Dockerfile testing and repair process in detail. After a Dockerfile is generated by \name (Fig.~\ref{fig:dockerfile-repair}\circled{1}), it needs to be tested in a safe environment.
Since an installation process may involve unknown code and may need admin permissions, a degree of risk is involved when arbitrarily running installation scripts.
As such, we choose to test the agent's installation process using Dockerfiles running in virtual machines, thus mitigating the side-effects of running potentially harmful or disruptive code (Fig.~\ref{fig:dockerfile-repair}\circled{2}).

Once the Dockerfile build process has finished, the resulting logs are analysed.
Installation is considered successful if tests are run, and at least one test passes.
Such a result is indicative of a successful installation as the tests running at all implies that dependency issues were not encountered during the setup of the test suite.
This is an imprecise measure of build success as, while requiring all tests to pass could lead to false negatives as tests could fail due to issues unrelated, such as API keys not being set, cases where different modules of a project have different requirements could incorrectly pass, resulting in false positives.
In the case that any tests passed, the process has finished and the agent has completed its task. If no tests passed or the installation process failed at an earlier step, \name starts the Dockerfile repair process.


The repair process starts with a new system message including both the Dockerfile the agent previously submitted, as well as the build logs of the failed installation process (Fig.~\ref{fig:dockerfile-repair}\circled{3}). The agent is subsequently instructed to explain what the error message means, and to identify the cause of the error.

Similarly to the previous steps, the agent is given access to the basic search functions. However, the agent is no longer encouraged to inspect only documentation files, but any files it considers to be relevant to the issue. The messages from the previous steps are disregarded, meaning this is effectively a new agent instance, with no knowledge of the previous stages.
This is done to reduce the context length of the messages being sent to the LLM, which can both improve its performance~\cite{liu2024lost} and also reduce inference cost.

After providing the explanation why the previous Dockerfile failed (Fig.~\ref{fig:dockerfile-repair}\circled{4}), the agent is instructed to suggest how this error could be fixed, as well as the repaired Dockerfile (Fig.~\ref{fig:dockerfile-repair}\circled{5}), which is then sent to the local virtual machine to be tested again. This process is repeated until \name produces a working Dockerfile, or it reaches a fixed number of repair attempts, at which point we consider the attempt to install the repository a failure. Due to the high time cost of additional repair attempts, we set the maximum number of repair attempts to two.

\begin{figure*}[ht!]
    \centering
        \begin{subfigure}[T]{0.32\textwidth}
            \centering
            \includegraphics[width=\linewidth]{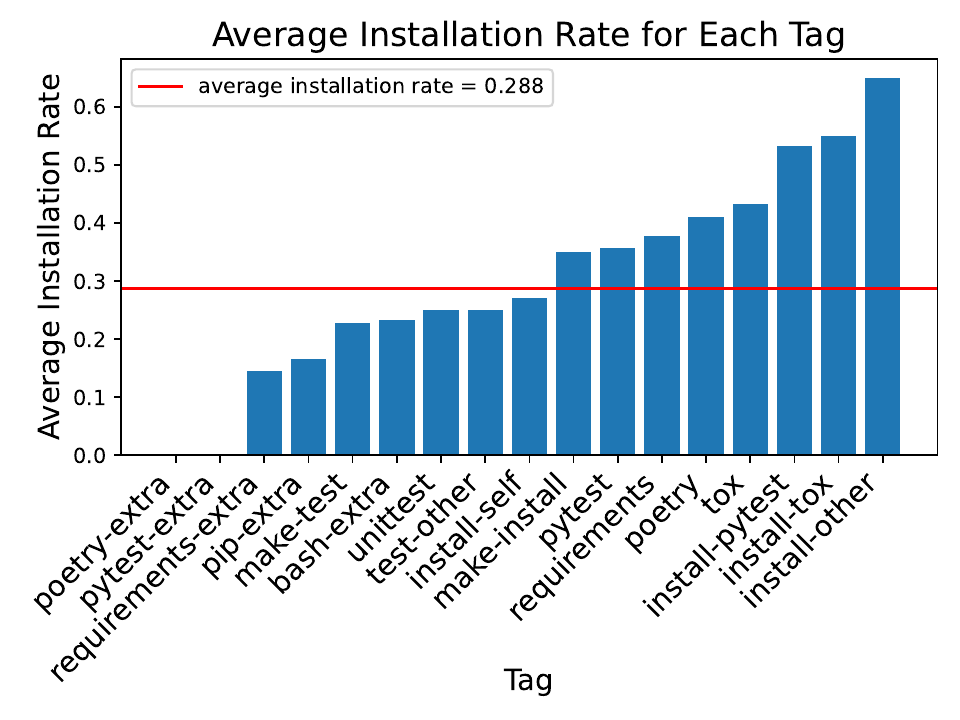}
            \vspace{-16pt}
            \caption{Average install rate for each tag}
            \label{fig:br-by-tag}
        \end{subfigure}
    \hfill
        \begin{subfigure}[T]{0.32\textwidth}
            \centering
            \includegraphics[width=\linewidth]{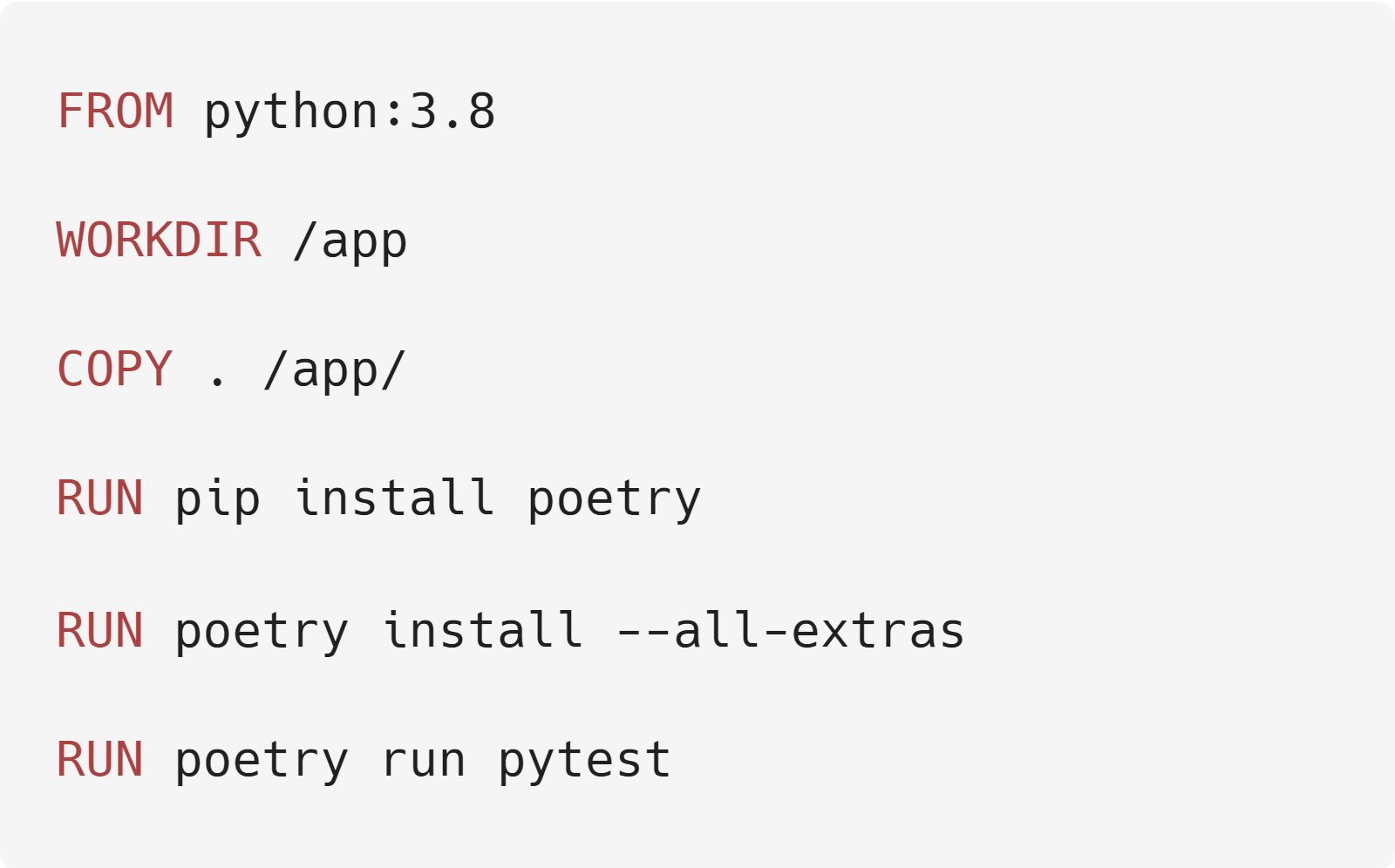}
            \vspace{5pt}
            \caption{Exemplar DockerFile for the NoneBot2 repository}
            \label{lst:nonebot2}
        \end{subfigure}
    \hfill
        \begin{subfigure}[T]{0.32\textwidth}
            \centering
            \includegraphics[width=\linewidth]{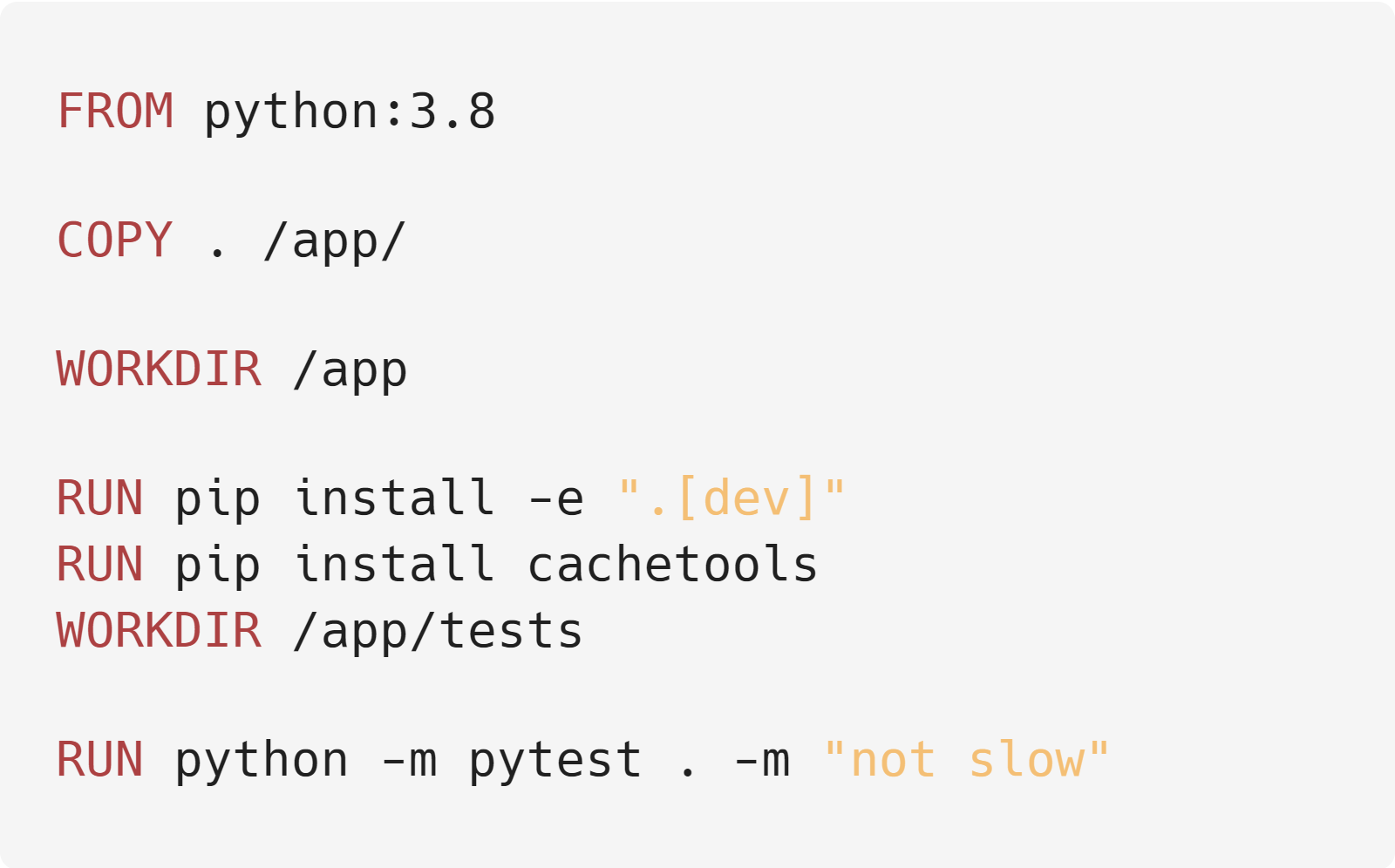}
            \vspace{5pt}
            \caption{Exemplar DockerFile for the Qlib repository}
            \label{lst:qlib}
        \end{subfigure}
    \caption{Identifying causes of un-installable repositories}
    \vspace{-15pt}
\end{figure*}

\section{Experimental Setup}
\label{sec:experimental-setup}

\subsection{Research Questions}
In the creation and evaluation of this agent, the following three research questions are posed:
\begin{itemize}
    \item \textbf{RQ1: To what extent is our agent able to install arbitrary Python repositories?}\\
        This question analyses our agent's performance on our benchmark dataset and aims to determine the circumstances under which our agent is and is not able to automatically install a repository.
    \item \textbf{RQ2: What factors affect an LLM agent's ability to successfully install a repository?}\\
        Understanding what factors could affect an agent's compatibility with a repository could prove invaluable in the case that the use of LLM agents for repository-level tasks become widespread.
        As such, we aim to identify the key characteristics of repositories that are easier and harder to automatically install.
    \item \textbf{RQ3: What challenges remain in designing agents for automatic installation and similar repository level tasks?}\\
        To aid future research in the pursuit of using LLM agents for environment management tasks,
        we outline the greatest challenges and limitations we encountered during the ideation and development of our agent.
\end{itemize}




\subsection{Metrics}
\label{sec:metrics}
We aim to study the relationship between the quality of documentations, and the performance of our agent. To estimate the quality of documentations, we propose two measures of document quality: \visibility and \informativity.

The \visibility of a repository's documentation intuitively measures how easy it is to find the install-relevant documents. For this, we count the total number of files and directories that need to be traversed to reach all install-relevant documents, and take its inverse:

$$\text{\visibility} = \frac{1}{\text{\#directories} + \text{\#files}}$$


The \informativity of a repository's documentation is, intuitively, a measure of how comprehensive the documentation is with respect to the actual installation process as defined by our ground truth Dockerfile. For this, we compute the proportion of lines of the ground truth Dockerfile that also appear in install-relevant documentations:

$$\text{\informativity} = \frac{|\text{dockerfile}\bigcap \text{documentation}|}{|\text{dockerfile}|}$$

Note that these two heuristic measures use the gathered ground truths; thus, their computation relies on manual inspection of the studied repositories. We use these measures in Section~\ref{sec:RQ1} to determine the relationship between the agent's success rate and the quality of the documentation, with the expectation that repositories with poorer quality, as estimated by these metrics, will be more difficult for the agent to install and consequently will have a lower average installation rate.

To measure installation performance, we use two other metrics: the recall from its documentation gathering step, and the successful installation rate of the Dockerfile generation step, across repeated runs. The recall is computed as follows: 

$$\text{recall} = \frac{|\text{install-relevant retrieved documents}|}{|\text{install-relevant documents}|}$$

Whereas the successful installation rate is simply the proportion of successful installations among the repeated runs. We apply \name to each repository 10 times.

\subsection{Implementation}
\label{sec:implementation}

All experiments have been conducted using the GPT4o-mini (\texttt{gpt-4o-mini-2024-07-18}) model; the Dockerfiles have been built in a virtual machine running Ubuntu 22.04.4 LTS and Docker 27.1.2. All Python scripts in the artiact were run on Python 3.10.12 and figures' coefficients are calculated using the \texttt{polyfit} method of NumPy 2.0.1

\section{Results} 
\label{sec:results}
Here we present the findings from our empirical evaluation of \name.

\begin{figure}[htbp]
    \vspace{-5pt}
    \centerline{\includegraphics[width=1\linewidth]{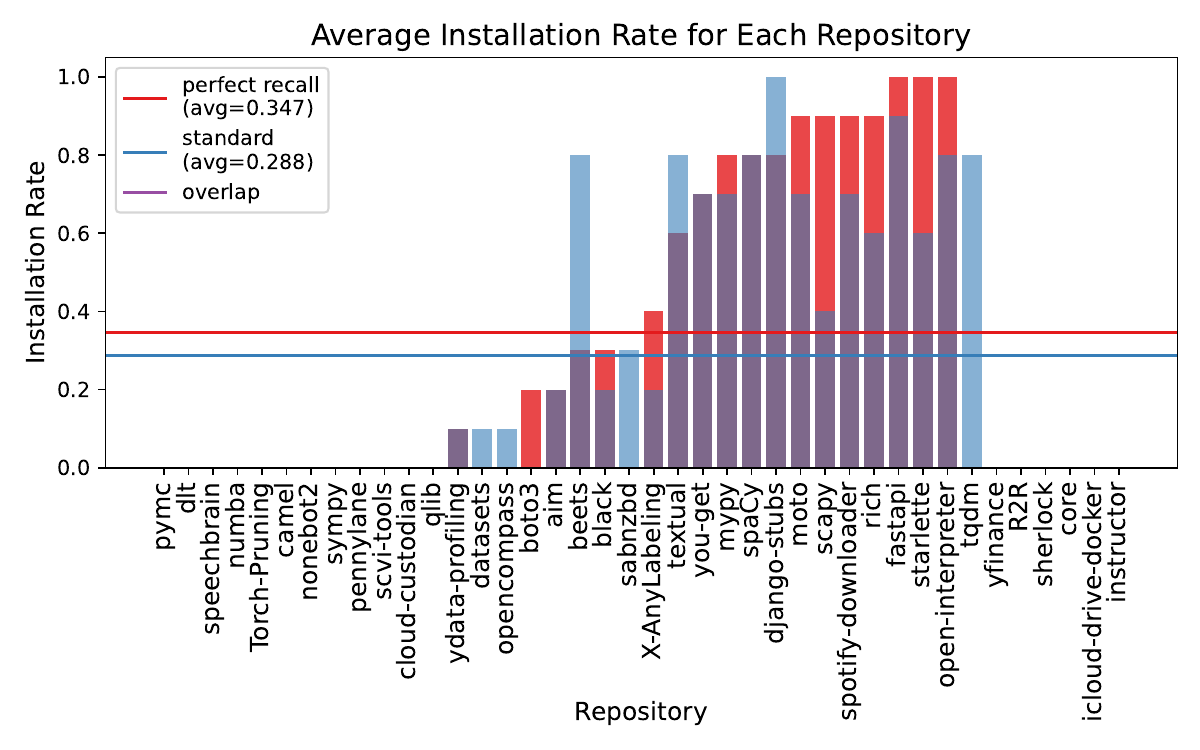}}
    \caption{Successful install rate for each repository, with and without perfect recall of relevant documents (purple bars represent the overlap between these two metrics).}
    \label{fig:install-rate}
    \vspace{-10pt}
\end{figure}

\begin{figure*}[!htbp]
    \centering
        \begin{subfigure}[T]{0.32\textwidth}
            \centerline{\includegraphics[width=\linewidth]{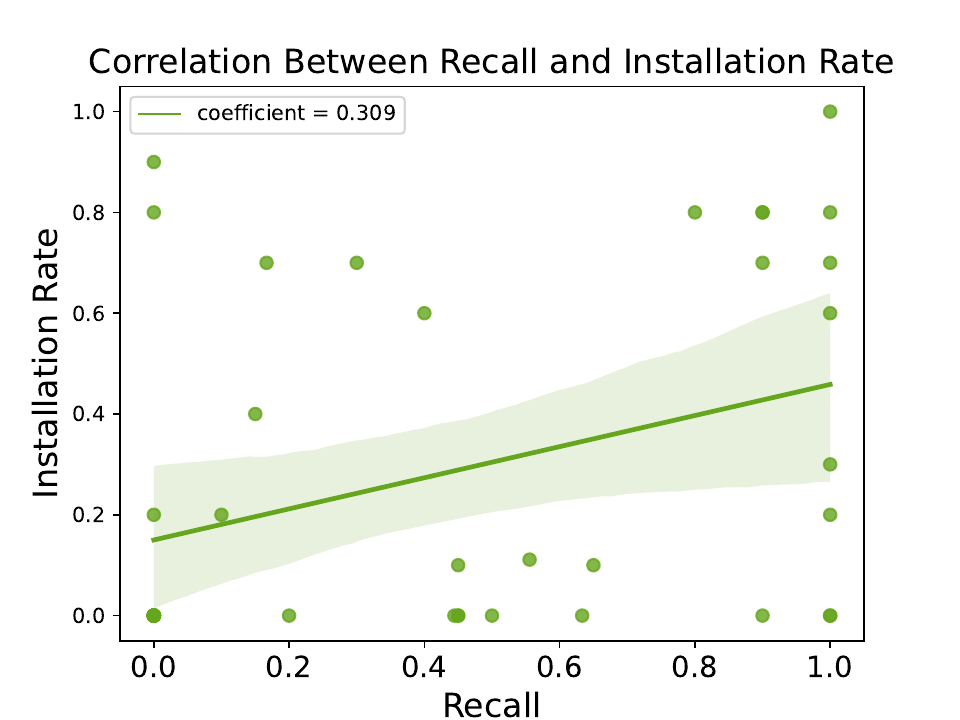}}
            \caption{Correlation between the documentation search recall and the average install rate}
            \label{fig:recall-br}
        \end{subfigure}
    \hfill
        \begin{subfigure}[T]{0.32\textwidth}
            \centerline{\includegraphics[width=\linewidth]{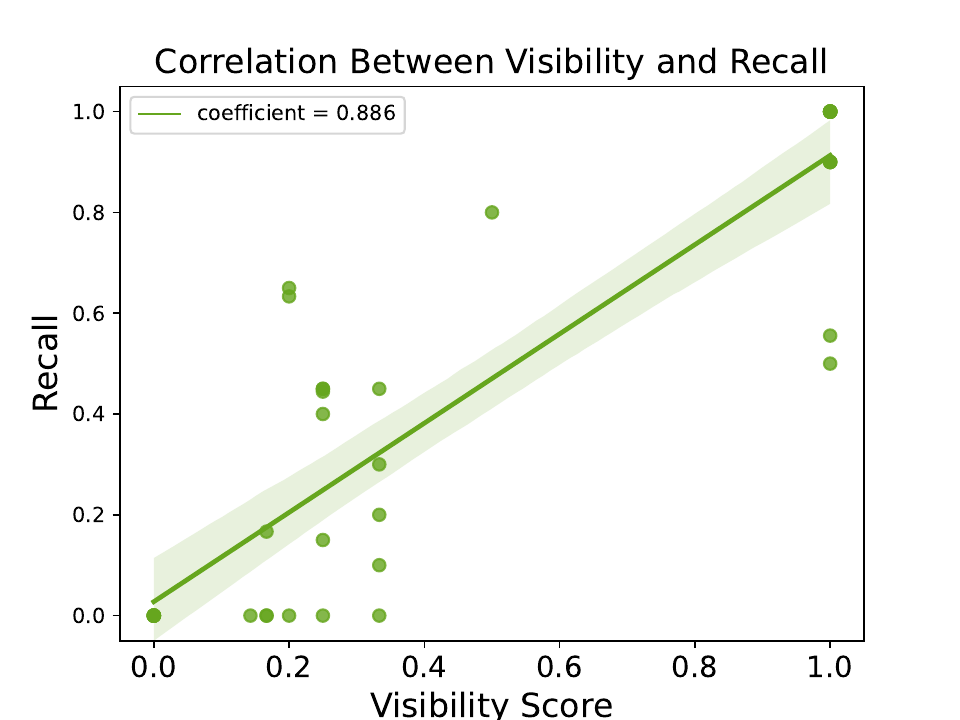}}
            \caption{Correlation between the recall of the documentation search step and \visibility}
            \label{fig:recall-vis}
        \end{subfigure}
    \hfill
        \begin{subfigure}[T]{0.32\textwidth}
            \centering
            \includegraphics[width=\linewidth]{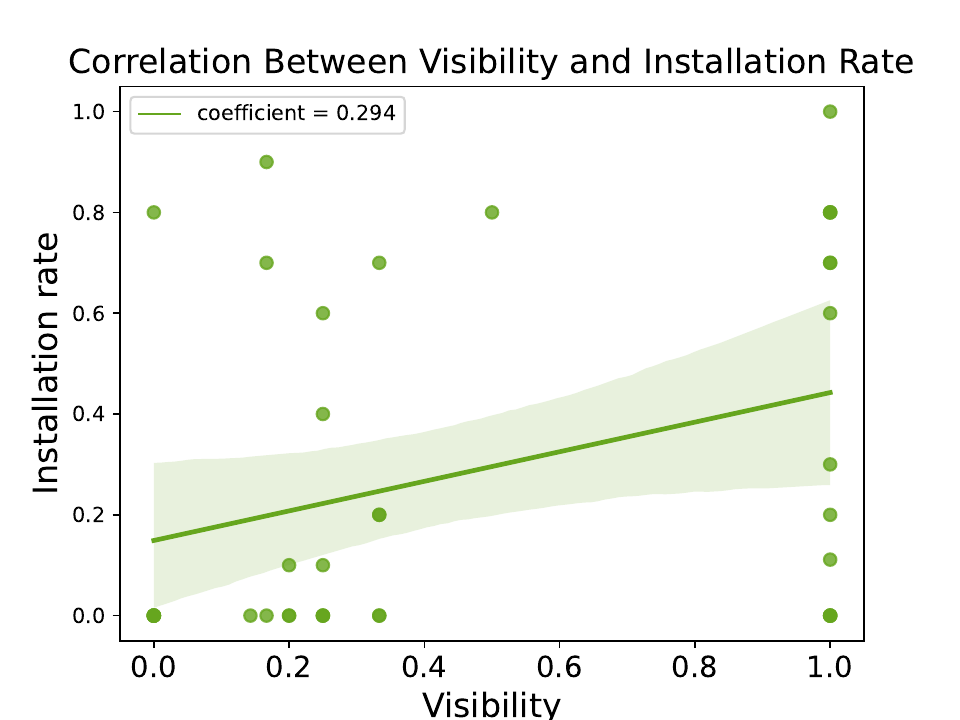}
            \caption{Correlation between \visibility and and average installation rate}
            \label{fig:vis-br}
        \end{subfigure}
    \caption{Evaluating the \visibility of a repository's documentation and its average installation rate}
    \vspace{-15pt}
\end{figure*}
\subsection{RQ1: To what extent is our agent able to install arbitrary Python repositories?} 
\label{sec:RQ1}

\subsubsection{Installation Success Rate}
Figure~\ref{fig:install-rate} shows the rates of successful installations achieved by \name for the studied repositories under two configurations.
The first configuration, \emph{standard} (blue bars in Figure~\ref{fig:install-rate}), is the standard configuration of \name including the documentation gathering step: the agent searches through the repository and select any items of documentation it deems to be install-relevant before attempting to write a Dockerfile.
The second configuration, \emph{perfect recall} (red bars in Figure~\ref{fig:install-rate}), is \name without the documentation gathering step: instead, we provide \name with all documents that we manually confirmed to be install-relevant.
The second configuration is included to isolate the installation success rate from the \visibility of the repository's documentation.
The seven rightmost repositories do not contain any relevant documentation in their repositories and consequently are excluded from the perfect recall configuration.
Of these seven repositories, only \texttt{tqdm}
was successfully built by the agent.

Under standard configuration, \name can successfully install 21 of 40 tested repositories at least once, with a 28.8\% average successful installation rate across all repositories. Under the perfect recall configuration, \name can install 18 of 34 tested repositories, with a 34.7\% average installation rate, a roughly 20\% increase. Of the 34 repositories tested in both configurations, 13 are never successfully built.

Note that there are several repositories whose installation rate is higher in the standard configuration than in perfect recall configuration. This is likely due to variance in the LLMs beahaviour (as only 10 attempts were made per repository), rather than the repository actually being easier to install with imperfect recall of install-relevant documentation.

\subsubsection{Causes of Failure}
In order to identify the limitations of the agent's capabilities, we consider the relationship between each of the 17 previously defined tags with the installation success rate of the repository in Figure~\ref{fig:br-by-tag}.
Of the tags present across all 40 repositories, there are only two which the agent is never successful in installing: \textbf{poetry-extra} and \textbf{pytest-extra};
example installation processes that include these tags can be seen in Figures~\ref{lst:nonebot2} and~\ref{lst:qlib} respectively.

Figure~\ref{lst:nonebot2} shows the exemplar Dockerfile written for to install and test the NoneBot2
repository.
In this example, the additional \texttt{--all-extras} argument needs to be added to the installation command as the repository's test suite tests modules with dependencies not installed by \texttt{poetry install} on its own.
Despite this, the \texttt{--all-extras} tag is not mentioned in the install-relevant documentation, resulting in the agent not being aware of this problem, even in the case when the documentation is provided.

Figure~\ref{lst:qlib} shows a Dockerfile for the \texttt{Qlib}
repository, and features the other un-installable tag, \textbf{pytest-extra}.
Here, two additions are made to the standard testing process. First, the user must run the test suite from within the \texttt{test} directory, and second, the test suite must be run with the \texttt{"not slow"} argument.
Neither of these additions were mentioned of the documentation for the repository, although, this is to be expected of the second requirement.
While first of these two additions is necessary for the test suite to run correctly, the second addition is due to our testing process, rather than the nature of the repository itself.
In order to prevent Dockerfile build processes from stalling and never completing, usually due to connection issues, we enforce a 30 minute time limit on the build process; any Dockerfile that takes over 30 minutes to build is interrupted and considered insufficient.
As such, it is common for repositories with larger test suites to be run with an additional command such as the aforementioned \texttt{"not slow"} to allow for the testing process to finish in time.
While important to guarantee that our experiments eventually finish, the requirement for test suites to finish within 30 minutes nonetheless is a detriment to our agent's installation rate, as without it, it could be the case that repositories such as \texttt{Qlib} could be successfully built.

\begin{figure*}
    \centering
        \begin{subfigure}[B]{0.32\textwidth}
            \centerline{\includegraphics[width=0.9\linewidth]{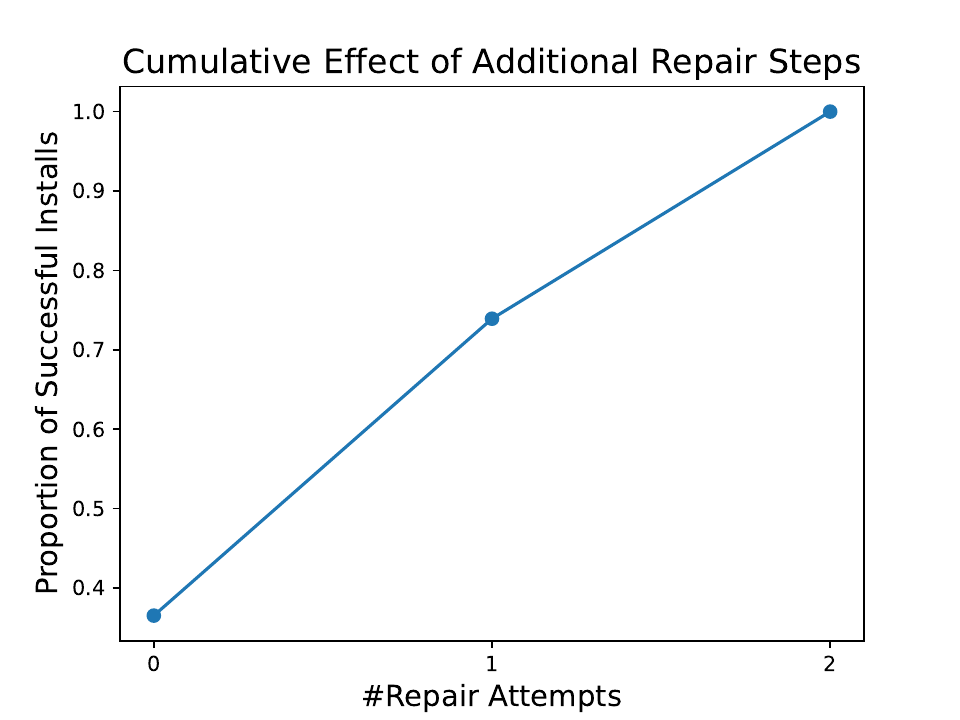}}
            \caption{Proportion of successful installations for each repair step (cumulative)}
            \label{fig:repair-step}
        \end{subfigure}
    \hfill
        \begin{subfigure}[B]{0.33\textwidth}
            \centerline{\includegraphics[width=0.9\linewidth]{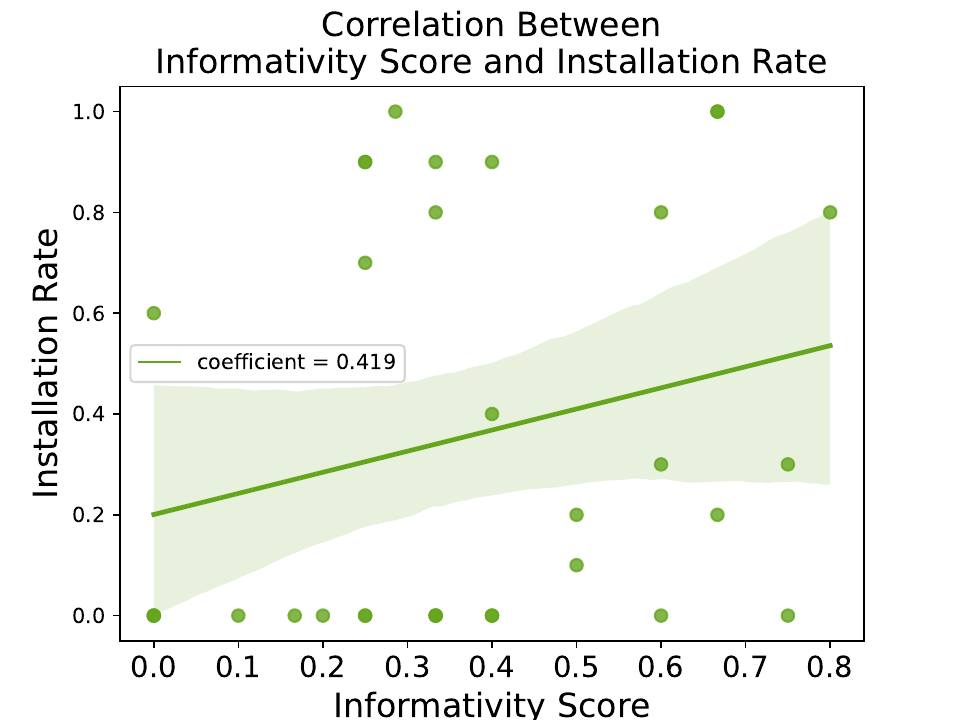}}
            \caption{Correlation between the \informativity score and the installation success rate}
            \label{fig:info-br}
        \end{subfigure}
    \hfill
        \begin{subfigure}[B]{0.33\textwidth}
            \centerline{\includegraphics[width=0.9\linewidth]{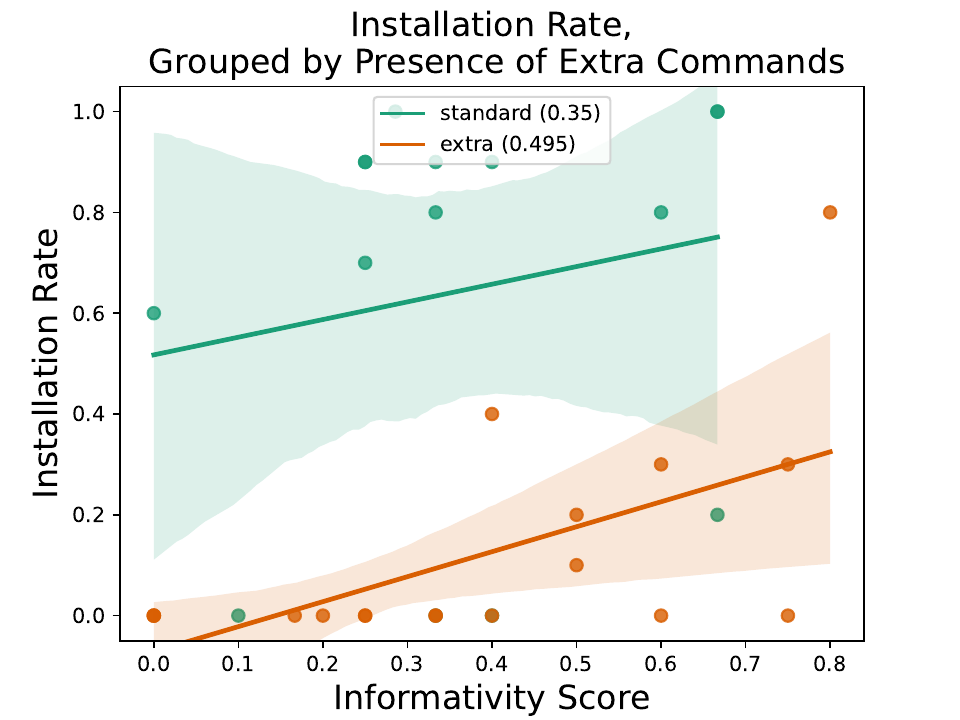}}
            \caption{Effect of extra commands on the installation rate of a repository}
            \label{fig:extra-tags}
        \end{subfigure}
    \caption{Factors affecting a repository's installation rate}
    \vspace{-15pt}
\end{figure*}

In both of these cases, the lines or additional arguments that are required for the installation process to succeed are not mentioned in the install-relevant documentation.
While these two tags are not present in all un-installable repositories, we were able to confirm that the presence of these two tags were the most common reason for failure of these repositories.

\begin{tcolorbox}[boxrule=0pt,frame hidden,sharp corners,enhanced,borderline north={1pt}{0pt}{black},borderline south={1pt}{0pt}{black},boxsep=2pt,left=2pt,right=2pt,top=2.5pt,bottom=2pt]
\textbf{Answer to RQ1:} \name can successfully install 21 out of 40 studied repositories, with the average installation success rate of 28.8\%. Most common reason for failed installation is that a successful installation process requires commands or arguments that are not mentioned in the relevant documentation.
\end{tcolorbox}

\subsection{RQ2: What factors affect an agent's ability to successfully install a repository?}
\label{sec:RQ2}

The first factor we consider is that of \name's own repair step. Figure~\ref{fig:repair-step} shows the proportion of successful installations that occur within each repair step. Only 36.5\% of successful installations (10.5\% of all installation attempts) occurred without the need for any repairs, increasing to 73.9\% after a single repair attempt (21.3\% of all attempts). These results indicate that further repair steps could potentially increase the installation rate, though this would come at the cost of increased installation time, as will be discussed in RQ3.

Figure~\ref{fig:recall-br} shows the relationship between the average recall of install-relevant documents and the installation rate of each repository.
The correlation is positive, confirming our intuition that the documents we marked as install-relevant do indeed assist the agent in successfully installing the repository. 
Figure~\ref{fig:recall-vis} shows a strong positive correlation between the average recall and the \visibility of a repository. 
Consequently, the correlation between \visibility and installation rate is similarly positive, shown in Figure~\ref{fig:vis-br}.
This consistency between the relationships of recall and \visibility with the agent's installation rate
leads us to conclude that the structure of a repository's documentation does have an effect on the ability of our agent to install the repository; documentation made of fewer files, stored within fewer subdirectories, will be more compatible with LLM-based agents.

The second of our two proposed document quality metrics, \informativity, also correlates with our agent's ability to install a repository.
Figure~\ref{fig:info-br} shows the correlation between the \informativity score of a repository
-- the number of lines in our exemplar Dockerfile that appear in the repository's install-relevant documentation -- and the average installation rate.
In order to isolate the effect of \informativity on the installation rate, the results shown here are from the experiment in which all install-relevant documents are given to the agent in place of the documentation gathering step.
Similarly to \visibility, the correlation between \informativity and installation rate is positive, indicating that the agent responds well to documentation that contains lines of code that appear in the necessary installation steps.
However, \informativity does not consider other features of documentation, such as natural language instructions. Consequently, it is unclear whether instructions other than lines of code are more or less effective in supporting the agent's installation attempts.


\begin{figure*}
    \centering
        \begin{subfigure}[T]{0.32\textwidth}
            \centering
            \includegraphics[width=\linewidth]{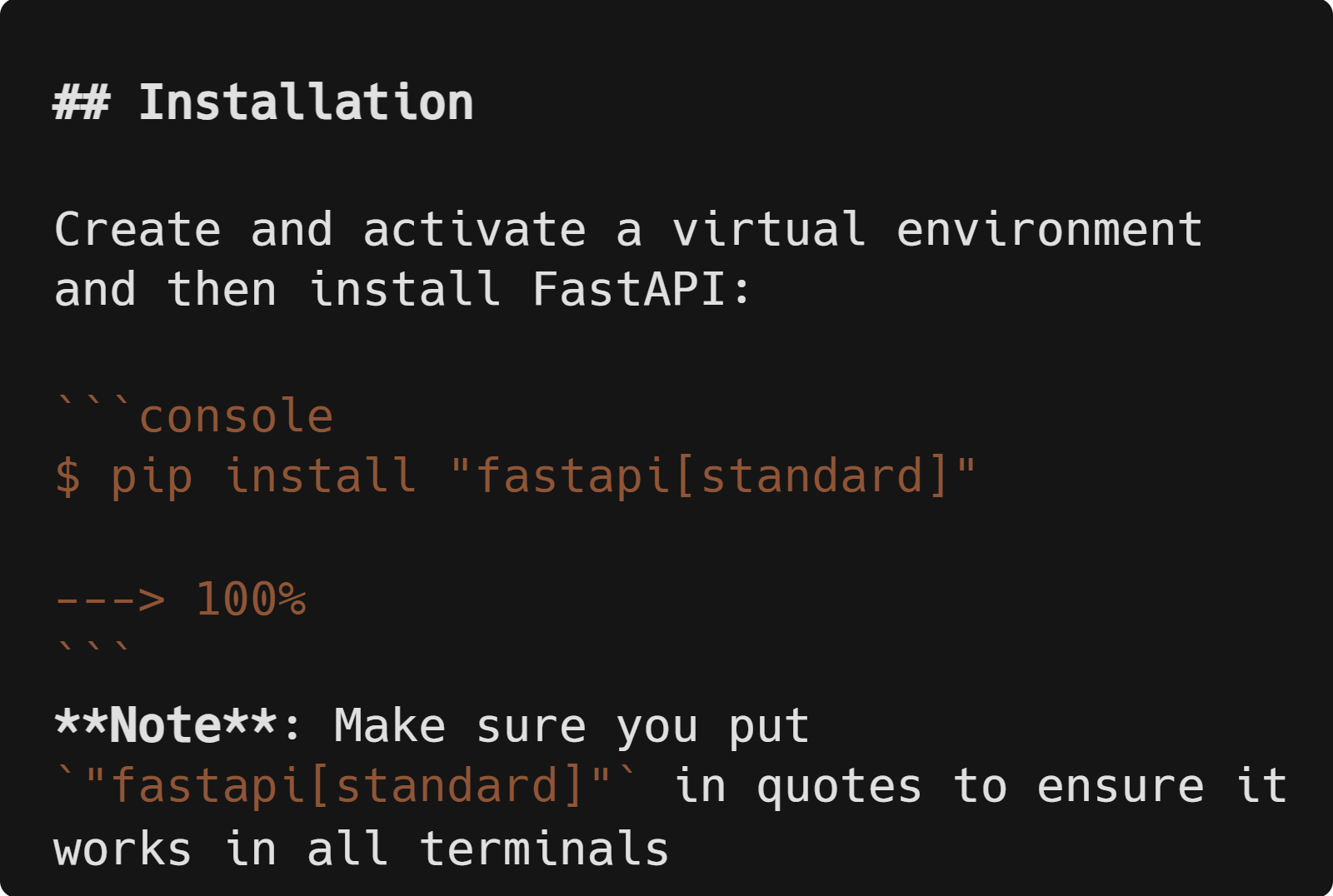}
            \caption{Example of documentation that is not relevant to setup of a development environment}
            \label{lst:fastapi-distractor}
        \end{subfigure}
    \hfill
        \begin{subfigure}[T]{0.32\textwidth}
            \centering
            \includegraphics[width=\linewidth]{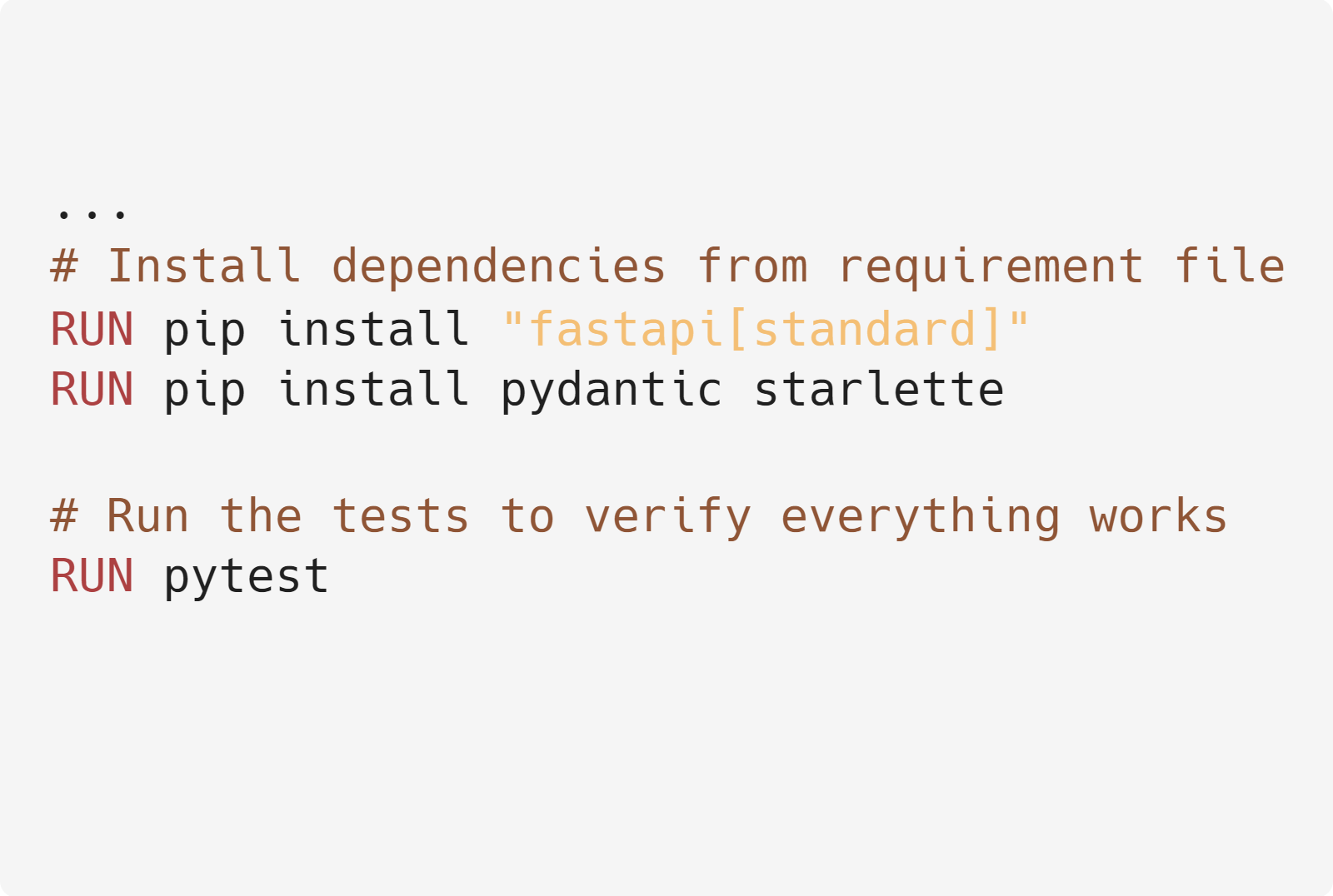}
            \caption{Incorrect Dockerfile generated for FastAPI}
            \label{lst:fastapi-bad-install}
        \end{subfigure}
    \hfill
        \begin{subfigure}[T]{0.32\textwidth}
            \centering
            \includegraphics[width=\linewidth]{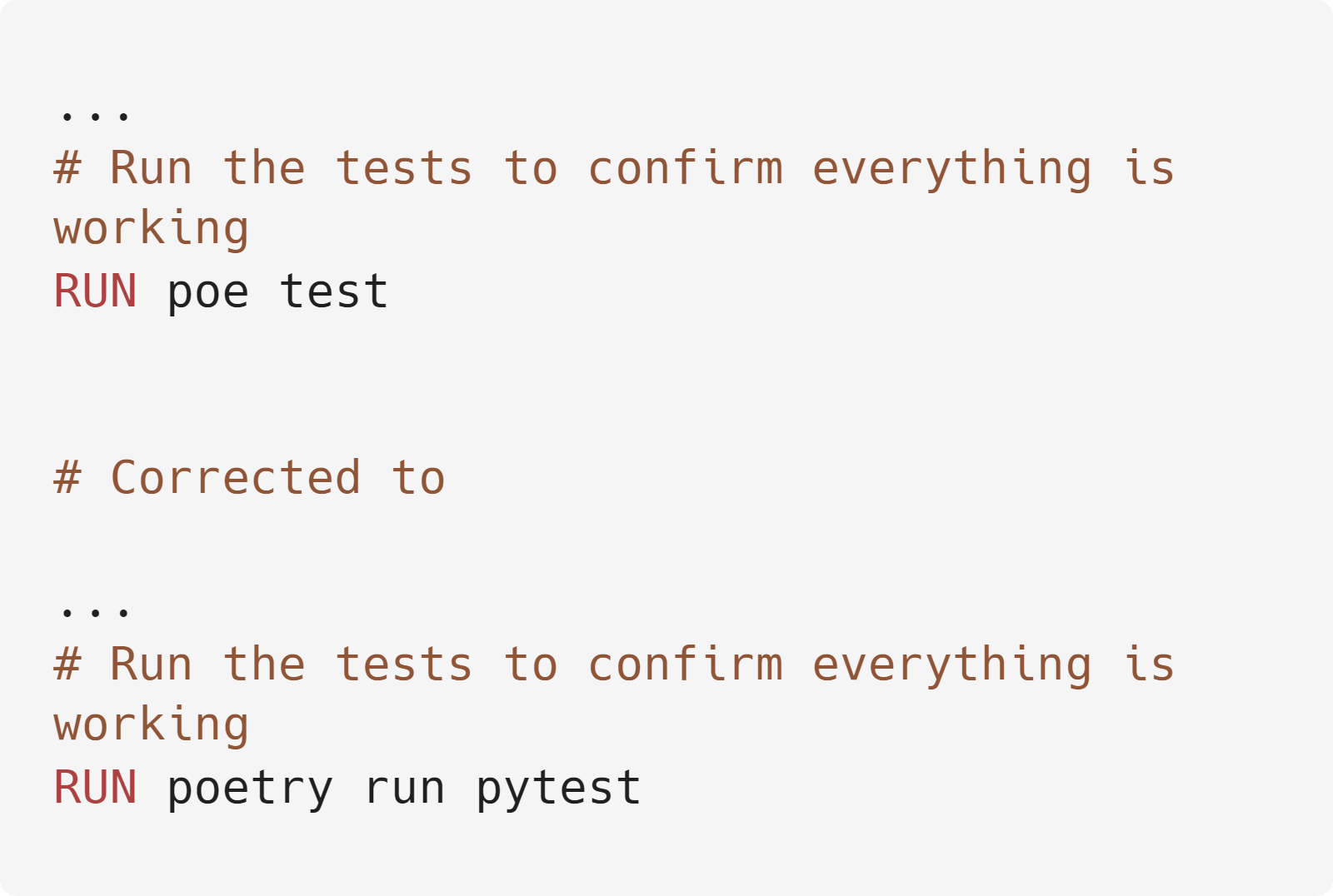}
            \caption{Incorrect Dockerfile generated for beets}
            \label{lst:beets-typo}
        \end{subfigure}
    \caption{Examples of distracting documentation and incorrect Dockerfiles}
    \vspace{-15pt}
\end{figure*}

Comparing the agent's installation rate only to our metrics of document quality - recall, \visibility, and \informativity -
would neglect the effect that the complexity of a repository's installation process has on the performance of the agent.
Unfortunately, the complexity of an installation process is hard to quantify,
and so too is the process of identifying and mitigating the effect of difficult to install repositories when evaluating our agent.
Instead, here we aim to do a qualitative estimation of the difficulty of a repository's installation process based on the types of tags it has been assigned with. 
Figure~\ref{fig:extra-tags} shows a clear difference in build rate of repositories that are tagged with \textbf{extra}
(\textbf{pip-extra}, \textbf{requirements-extra}, \textbf{poetry-extra}, \textbf{pytest-extra} or \textbf{bash-extra}) and those that are not.
It can therefore be inferred that, while the agent may be familiar enough with tools such as \texttt{pip}\footnote{https://pypi.org/}, Poetry or PyTest to perform basic tasks, the inclusion of additional complexity, such as the need of specific additional flags or installing requirements from multiple files
greatly reduces the agent's ability to write a correct Dockerfile to install the repository. We note that this increased difficulty can be curbed to some extent through the addition of thorough documentation on the installation process, as demonstrated by the positive correlation between \informativity and the installation rate observed from those repositories tagged with \textbf{extra}.


\begin{tcolorbox}[boxrule=0pt,frame hidden,sharp corners,enhanced,borderline north={1pt}{0pt}{black},borderline south={1pt}{0pt}{black},boxsep=2pt,left=2pt,right=2pt,top=2.5pt,bottom=2pt]
    \textbf{Answer to RQ2:} The inclusion of additional complexity to a repository's installation process has the greatest impact on the installation rate, though this effect can be mitigated through clear documentation structure and the use of code examples in the installation instructions.
\end{tcolorbox}

\subsection{RQ3: What challenges remain in designing agents for automatic installation and similar repository level tasks?}
\label{sec:RQ3}

Through the process of creating and evaluating this agent, we faced four main challenges which limit both the effectiveness and practical feasibility of our agent.







\subsubsection{Identifying install-relevant documentation}

As reported in Section~\ref{sec:doc_structure}, install-relevant documentation  can be found in various locations within a repository. \name shows recall of zero for 11 out of 40 repositories, despite there being only six repositories in the dataset without any install-relevant documentation. The lack of consistent structure in documentations of open source Python repositories affects the ability of LLM-based agents to identify install-relevant documents. RQ1 and RQ2 show that the failure to retrieve install-relevant documents
negatively affects the installation rate of the repository substantially.

An example of how irrelevant information can lead to incorrect Dockerfile can be found in the `Installation' section of the the \texttt{README.md} file of \texttt{FastAPI}, shown in Figure~\ref{lst:fastapi-distractor}.
The section provides installation instructions for users intending to use \texttt{FastAPI} to develop their own tools, rather than those who want to work on the \texttt{FastAPI} code itself. While it may be clear to a human developer whether this section is relevant to them, our evaluation of \name shows that LLM-based agents may be susceptible to misunderstanding the intention of this document, which is shown in Figure~\ref{lst:fastapi-bad-install}, where the generated Dockerfile erroneously follows the instructions of the irrelevant section, rather than correctly installing the dependencies of the project from the \texttt{requirements.txt} file.
In this case, the installation failed when trying to run PyTest, as Pytest is not included in \texttt{"fastapi[standard]"}, \texttt{pydantic} or \texttt{starlette}.

\subsubsection{Writing valid Dockerfiles} 
\label{sec:valid-files}

\name often fails to write a correct Dockerfile, not because it misunderstands the installation instructions, but simply because it made mistakes while writing the Dockerfile itself. Figure~\ref{lst:beets-typo} gives an example, i.e., a typo made by \name in one of the generated Dockerfiles for the \texttt{beets} repository.
As shown, this typo was identified and fixed by the agent during the repair step.

While we maintain that attempting to install the repositories inside a Docker container is necessary to mitigate the effect of any potentially harmful code,
installation and testing in the manner we instruct the agent to do is not the usual purpose of Docker, and we suspect that this unfamiliarity with writing installation scripts in a Dockerfile could lead to additional mistakes.
A previous work~\cite{liang2024gpt4replicateempiricalsoftware} has encountered a similar issue of an LLM agent being prone to low-level mistakes, despite seeming to have a high level understanding of the given task.

\subsubsection{Cost}
In previous works, repository level techniques that search through the contents of a repository delegate this search process to static analysis tools~\cite{wang2024teaching} or a non-LLM based search technique, such as BM25~\cite{tao2024magis}, and then include the search results in a prompt to the LLM.
In order to emulate the behaviour of a human developer, our search process is controlled by the LLM itself.
This means that the number of tokens used by the agent is largely unconstrained, resulting in considerably higher usage than related work.
The additional cost can vary greatly depending on the contents of the target repository; extensive documentations or ambiguous naming can lead to the agent searching through more files than necessary, and thus consuming more tokens~\cite{liu2024lost,li2024long,zhou2023thread}, resulting in higher cost.
Aside from the financial cost incurred, the time taken by the agent to attempt an installation of a repository can vary considerably depending on the size of its dependencies, its test suite and of the repository itself; an installation attempt takes 501 seconds on average with \name, with the longest run taking almost 80 minutes over the course of Dockerfile building attempts.

\subsubsection{Oracle problem}

While we choose to use a repository's test suite for the oracle, this is not an ideal solution for various reasons. It is not a trivial task for any agent to run tests for an arbitrary project, as Python supports several different methods of testing. Any agent aiming to run tests for an arbitrary Python project should first correctly identify the method of test execution, on top of correct identification of the installation method, adding an extra layer of complexity to the overall task.
Additionally, it is also possible that a repository does not contain any test, making it impossible for our current design of \name to attempt installation.

Figure~\ref{fig:br-by-tag} shows that all testing related tags other than \textbf{pytest} contribute to a lower installation rate. In the case that these alternative testing methods are the cause of this reduced installation rate, using an oracle other than running the test suite would allow us to ignore the effect of these tags, as the agent's performance would no longer be dependant on its ability to run tests.

\begin{tcolorbox}[boxrule=0pt,frame hidden,sharp corners,enhanced,borderline north={1pt}{0pt}{black},borderline south={1pt}{0pt}{black},boxsep=2pt,left=2pt,right=2pt,top=2.5pt,bottom=2pt]
    \textbf{Answer to RQ3:} There are several limitations faced when using an LLM based agent to automatically install Python repositories, with the most critical being finding an oracle that is both generalisable and accurate.
\end{tcolorbox}

\section{Related Work} 
\label{sec:related-work}

\subsection{Repository-Level Tasks}
Bairi et al.~\cite{bairi2024codeplan} define an LLM-driven repository-level coding task as one that requires a series of edits to be made to the state of a code base until some oracle is satisfied.
Previous works have proposed the use of LLMs to perform tasks that satisfy this definition of repository-level coding tasks~\cite{tao2024magis,wang2024opendevin},
and are typically evaluated on benchmarks such as SWE-bench~\cite{jimenez2023swe}.
SWE-bench tasks a language model to edit a codebase to address a description of an issue.
These edits are then assessed using a test suite, as well as through comparison with a human written pull request that resolves the issue.

This task of issue resolution posed in SWE-bench is a clear example of a repository-level coding task as described by Bairi et al.~\cite{bairi2024codeplan}. However, the definitionof Bairi et al. is rather narrow, and does not consider other examples of repository-level software engineering that has been proposed in recent works. AutoFL~\cite{kang2024quantitative} is a fault localization technique that makes use of function calls to search through the repository and find potentially erroneous lines. It fits our definition of repository-level task, as it makes use of information across mlutiple files in a repository, but not that of Bairi et al., as AutoFL is a debugging technique and does not make any edits. 
Other works propose techniques that consider multiple files in a repository, but are only tasked with generating code for a single function~\cite{wang2024teaching}. In tasks such as this, the scope of the changes is limited to a single file, but the agent is nonetheless able to access files across the whole repository.

\subsection{Automated installation}
To the best of our knowledge, no previous literature has addressed the challenge of automatic installation of arbitrary repositories, although there are some works that have addressed similar tasks.
A recent work by Guerrero, Corcho and Garjio~\cite{utrilla2024automated} proposes PlanStep, a methodology to extract structured installation instructions from README files of research software projects through the use of LLMs.
While this project is very similar to the initial search step of our proposed agent, PlanStep's goal is to identify all possible methods of installation, rather than performing the installation itself.

Cognition AI's Devin\footnote{https://www.cognition.ai/blog/introducing-devin} project, which claims to be an `AI software engineer', does seem capable of automatically installing an open-source repository;
Devin has been demonstrated cloning a repository and installing its dependencies, given only the GitHub URL of the repository.
Unfortunately Devin is not publicly available at the time of writing, so its exact capabilities are currently unclear.

\subsection{Documentation analysis with LLMs}
In a recent survey~\cite{hou2023large} on the use of LLMs for software engineering tasks, while hundreds of papers on LLMs for software engineering were identified,
none used documentation as input to perform some task, and only one addressed the task of evaluating the quality of a code-base's documentation.
Furthermore, the work in question\cite{khan2021automatic} by Khan et al. specifically focus on detecting API Documentation smells, rather than high level documentation in a repository,
and as such is not relevant to the technique we propose.

Liang et al.~\cite{liang2024gpt4replicateempiricalsoftware} investigates the use of another form of documentation,
empirical software engineering papers, with the goal of replicating their research methodologies and results.
A key finding of this paper was that, when writing code to replicate the contents of these papers,  the LLM they used (GPT-4) "is correct in its high-level structure, but can contain errors in its lower-level implementation".
While neither the input nor the output of this task is directly comparable to the automatic installation of python repositories, we experienced a similar problem with the LLM that we use, GPT-4o-mini, as we discussed in Section~\ref{sec:valid-files}.

\section{Threats to Validity}
\label{sec:validity}
\emph{Threats to internal validity} are challenges to the findings of the paper. Due to the stochastic nature of LLMs, the behaviour and thus performance of our agent can vary between runs. To mitigate this randomness we repeat our experiments 10 times, and report the average scores over these 10 runs. For reproducibility, we make both our implementation and the messages generated in our experiments available for scrutiny.

\emph{Threats to extrenal validity} concern whether the reported findings may generalise to other results. The design of our agent allows for it to be applied to repositories not contained on our dataset, although the calculation of metrics such as recall would require additional manual inspection. We also design our agent to be agnostic to the LLM used to control it, allowing for experimentation with different models.

\emph{Threats to construct validity} concerns whether the measurements are actually based on the properties we are interested in. The quality of installation documentation, as well as their usability in terms of repository organization, is a highly abstract property and can be subjective. We aim to define clear and transparent heuristic measures for these with \visibility and \informativity. Other metrics such as recall and installation rate are both intuitive and straightforward.

\section{Conclusion} 
\label{sec:conclusion}

This paper studies a novel repository level task for LLM-based agents, namely automated installation of arbitrary repositories. In order to provide insight into this task, we make three contributions.
First, we created a dataset of 40 open source Python repositories for evaluating the effectiveness of repository level agents' understanding of documentation, as well as their ability to correctly install a repository.
Second, we present \name, an easily adaptable design for an LLM-based agent that is able to autonomously inspect the contents of a repository and recover items of documentation relevant to its task. An empirical evaluation of \name using our dataset shows that it can install 21 out of the 40 repositories, as well as a clear correlation between the structure and content of a repository's documentation and the performance of our agent. 
Finally, we report the challenges faced when developing an agent for repository-level agents for documentation,as well as suggestions to overcome these challenges for future developers. We hope that our dataset and empirical insights can contribute to future work on environment management tasks such as automated installation. 

\section*{Acknowledgement}
This work was supported by the National Research Foundation of Korea (NRF) funded by the Korean Government MSIT (RS-2023-00208998), and the Engineering Research Center Program funded by the Korean Government MSIT (RS-2021-NR060080).

\balance
\bibliographystyle{IEEEtran}
\bibliography{ref}

\end{document}